\documentclass[conference]{IEEEtran}
\IEEEoverridecommandlockouts

\usepackage{cite}
\usepackage{amsmath,amssymb,amsfonts}
\usepackage{algorithm}
\usepackage{algorithmic}
\usepackage{graphicx}
\usepackage{textcomp}
\usepackage{xcolor}
\usepackage[normalem]{ulem}
\usepackage{booktabs}
\usepackage{tikz}
\usepackage{subcaption}
\usepackage{fancyhdr}
\usetikzlibrary{shapes, arrows, positioning, patterns, shadows, calc}

\definecolor{customGray}{HTML}{A6A6A6}
\definecolor{customBlue}{HTML}{4E79A7}
\definecolor{customOrange}{HTML}{F28E2B}
\definecolor{customGreen}{HTML}{59A14F}
\newcommand{\rev}[1]{#1}

\usepackage[english]{babel}
\def\BibTeX{{\rm B\kern-.05em{\sc i\kern-.025em b}\kern-.08em
    T\kern-.1667em\lower.7ex\hbox{E}\kern-.125emX}}

\begin{document}

\title{An AI-Driven Framework for Energy-Efficient Environmental Monitoring in Smart Cities Using Edge Intelligence}

\author{
    \begin{tabular}{ccc}
        \begin{minipage}[t]{0.3\textwidth}
            \centering
            1\textsuperscript{st} Yichen Liu\\
            \textit{Independent Researcher}\\
            Mercer Island, WA, USA\\
            yil160@ucsd.edu \\
            ORCID: 0000-0003-3804-2970
        \end{minipage} &
        \begin{minipage}[t]{0.3\textwidth}
            \centering
            2\textsuperscript{nd} Imam Akintomiwa Akinlade\\
            \textit{Harvard Business School}\\
            Boston, MA, USA\\
            iakinlade@mba2023.hbs.edu\\
            ORCID: 0009-0005-1207-1652
        \end{minipage} &
        \begin{minipage}[t]{0.3\textwidth}
            \centering
            3\textsuperscript{rd} Xiaochong Jiang\\
            \textit{Independent Researcher}\\
            Seattle, WA, USA\\
            jiang.xiaoc@northeastern.edu\\
            ORCID: 0009-0008-0586-5851
        \end{minipage}
    \end{tabular}
    \\[4ex] 
    \begin{tabular}{cc}
        \begin{minipage}[t]{0.3\textwidth}
            \centering
            4\textsuperscript{th} Wenting Yang\\
            \textit{Independent Researcher}\\
            San Diego, CA, USA\\
            wey023@ucsd.edu\\
            ORCID: 0009-0009-3525-9227
        \end{minipage} &
        \begin{minipage}[t]{0.3\textwidth}
            \centering
            5\textsuperscript{th} Shiqi Yang\\
            \textit{Independent Researcher}\\
            New York, NY, USA\\
            sy3506@nyu.edu\\
            ORCID: 0009-0005-8846-7266
        \end{minipage}
    \end{tabular}
}

\maketitle

\thispagestyle{fancy}

\begin{abstract}
Environmental monitoring is a crucial component of the smart city infrastructure. It enables informed decision making which enhances sustainability, public health and urban planning. However, the large-scale deployments of the smart sensors have raised concerns on excessive energy consumption and redundant data collection as well as limited sensor lifespan. To resolve these issues, we present an AI-driven framework for energy-efficient environmental monitoring in smart cities utilizing edge intelligence. Our proposed framework leverages TinyML-enabled edge devices and context-aware adaptive decision-making in order to dynamically activate the sensors based on the spatiotemporal conditions, environmental statistics and energy constraints. The sensors will be dynamically activated based on a utility function that takes in factors such as real-time environmental conditions, sensor location, and remaining battery lifespan. Our framework will reduce unnecessary sensing and communication while maintaining high coverage for monitoring. We introduce a hierarchical Edge Intelligence architecture to support deployments in city-wide scales. We conducted evaluation using a city-scale simulation driven by real multi-sensor environmental traces, which demonstrates that the proposed mechanism significantly reduces energy consumption and extends sensor lifespan when compared to static, periodic, and UCB-based adaptive sensing strategies. The results highlight the potential of edge intelligence and adaptive AI techniques for building sustainable and efficient smart city monitoring systems.
\end{abstract}


\section{Introduction}
\label{sec:introduction}
Rapid urbanization has expedited the adoption of smart city technologies that intend to promote sustainability, health benefits, and improved quality of human life in urban areas \cite{b1}. Among the adopted smart city technologies, environmental monitoring systems provide real-time observations of various environmental parameters that affect the quality of life in urban areas. The use of large-scale sensor networks seeks to provide urban authorities with a better understanding of various environmental dynamics in order to proactively address emerging issues that affect urban areas, such as pollution events, urban heat island effects, and climate-related events \cite{b2}.

Despite their importance in smart city initiatives, various environmental monitoring systems face serious challenges in their operations and scalability in smart cities. For instance, most sensor networks deployed in various smart cities use static or periodic sensing paradigms where the sensor nodes operate continuously or periodically without considering the prevailing environmental conditions. Although static or periodic sensing paradigms are simple to implement in smart cities, they have shown serious shortcomings in terms of data redundancy, increased energy consumption, and reduced lifetimes of the sensor nodes in smart cities, particularly in cases where the sensor nodes are powered by batteries or have limited power supply \cite{b3}. These factors will impact the long-term deployment of the smart city monitoring systems and increase the overall cost of maintenance \cite{b4}.

Recently, significant advancements have been made in both areas of edge computing and Tiny Machine Learning (TinyML), which created new opportunities to increase the efficiency of IoT systems \cite{b5}. TinyML facilitates the execution of machine learning inferences on microcontrollers that require ultra-low power, which allows sensor devices to process data and make decisions locally \cite{b6}. This reduces the need to send data to a centralized server, and in turn, resulting in a reduction in communication costs, latency, as well as power consumption. However, most existing TinyML-based solutions have focused on the execution of local inferences, for example, anomaly detection and signal classification, in which scenarios the sensor activation strategies remain largely static.

However, the real-world setting in cities, especially, is naturally non-stationary and dynamic, with the conditions in the surroundings changing in terms of time and space. This, in turn, indicates that the informativeness of the sensor data will vary through time and space, and the sensor activation strategy will need to adapt to the context, including the locations, time of day, recent trends in the surroundings, and the availability of resources \cite{b7}. This has led to the need to design intelligent sensor activation strategies that can adaptively determine the most effective allocation of sensor resources in order to maximize the monitoring efficiency under the constraint of energy consumption \cite{b8}.

In this paper, we present a novel AI-based framework for energy-efficient environmental monitoring in smart cities based on the concept of edge intelligence. In the proposed framework, TinyML-based sensor nodes and an edge-intelligence-based decision-making mechanism are leveraged to control the activation of environmental sensors according to the spatiotemporal context and energy factors. Instead of maintaining the activation of all sensors at all times, only the most informative sensors are activated at each monitoring interval, thereby reducing redundant sensing and enhancing the overall lifetime of the network. Moreover, a novel Hierarchical Edge Intelligence architecture is proposed to support the deployment of the proposed framework in city-scale sensor networks.

The effectiveness of the proposed framework is evaluated by simulation-based experiments, where we compare our method with static and periodic sensing approaches. We can see through the results that the proposed framework significantly minimizes the energy consumption while maintaining high levels of environmental monitoring coverage, which makes it a viable and sustainable solution for smart cities.

The remainder of the paper is organized as follows. Section II discusses the related work regarding the concepts of smart city environmental monitoring, edge intelligence, and adaptive sensing strategies. Section III describes the proposed framework and the methodology. Section IV discusses the experimental evaluation. Finally, Section V provides the conclusion and future research directions.

\begin{figure*}[t!]
\centering
\includegraphics[width=\textwidth]{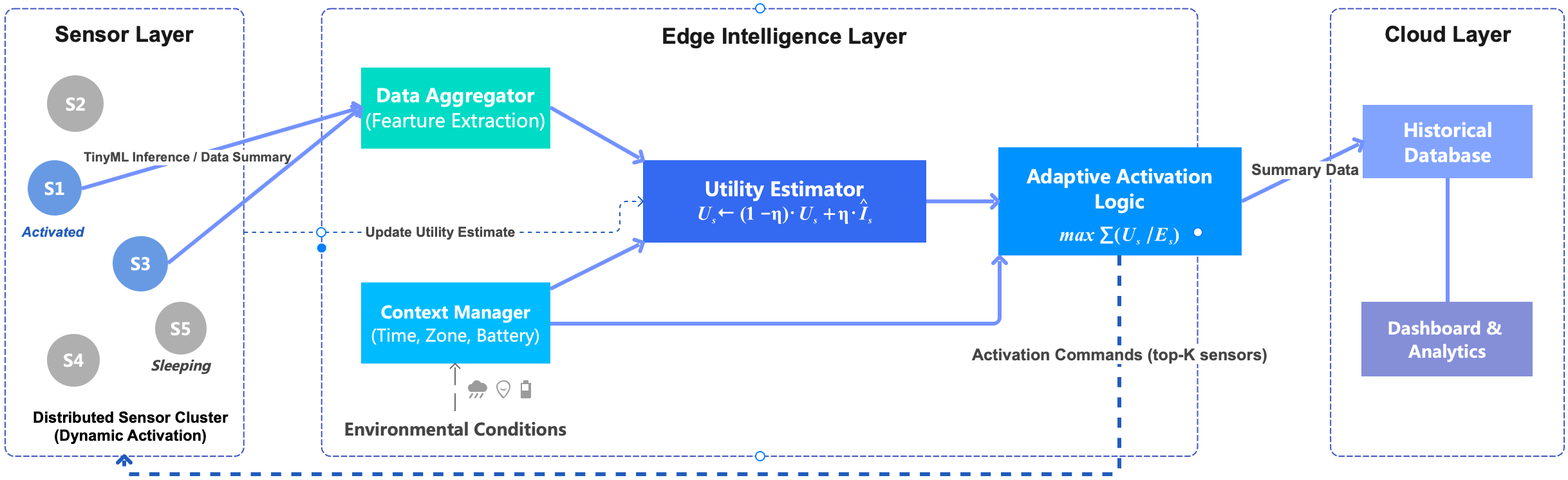}
\caption{Comprehensive system architecture diagram illustrating the interaction between the sensor layer, Edge Intelligence Layer, and cloud monitoring layer. The edge layer orchestrates sensor activation based on real-time context and energy constraints.}
\label{fig:system_diagram}
\end{figure*}

\section{Related Work}
\label{sec:related_work}
Energy-efficient monitoring systems lie at the intersection of wireless sensor networks (WSNs), edge computing, and adaptive control. This section reviews key advancements in these areas and identifies the gap addressed by our framework.

\subsection{IoT-Based Environmental Monitoring}
IoT-based urban sensing has been widely studied for city management. Most conventional architectures follow a `sense and send' model, in which nodes periodically transmit data to a centralized cloud infrastructure \cite{b1, b9}. Prior work improved air-quality monitoring and coverage, but still faces calibration, redundancy, communication delay, and battery-drain issues \cite{b10, b11, b3, b12}. Compressive sensing can reduce data size \cite{b13}, but it does not directly minimize sensing energy at the source.

\subsection{Edge Intelligence and TinyML}
To address the bottlenecks of cloud-centric systems, the paradigm has shifted toward edge computing \cite{b5}. TinyML further enables inference on resource-constrained microcontrollers \cite{b6}. Prior work has shown the feasibility of efficient edge models and on-device environmental inference \cite{b14, b15, b16}. However, most existing TinyML applications focus on optimizing the \textit{inference model} itself rather than using inference outputs to dynamically control sensing hardware or duty cycles.

\subsection{Adaptive and Context-Aware Sensing}
Adaptive sensing strategies seek to conserve energy by adjusting sensing behavior to changing conditions. Early methods used threshold-based sleep scheduling \cite{b17}, while later work exploited spatiotemporal correlations and decentralized coordination \cite{b18}. More recently, machine learning and reinforcement learning have been explored for adaptive sensing \cite{b8, b19}, and context-aware online learning has also been applied in related IoT domains \cite{b20}. Although promising, many of these methods remain too computationally heavy for extreme edge devices.

UCB-based online learning has also been explored for adaptive sensor selection in multi-sensing wireless sensor networks. In particular, Ghosh et al. developed a learning-based adaptive sensor selection framework that uses an upper-confidence-bound strategy to balance sensing quality and energy consumption while predicting inactive measurements from correlated active sensors \cite{b21}. This line of work confirms that lightweight bandit-based selection can serve as a strong adaptive baseline. However, these existing UCB-based methods primarily focus on node-level selection and do not explicitly incorporate hierarchical edge coordination, spatiotemporal context-aware utility estimation, or TinyML-generated local feedback for city-scale smart-city monitoring.

\subsection{Summary and Research Gap}
Existing systems are often either simple but inefficient, or adaptive but computationally expensive. There is still a lack of a lightweight framework that uses TinyML feedback to support real-time, energy-aware sensor activation without continuous cloud dependence. This paper addresses that gap through lightweight inference and adaptive utility-based selection.

\section{Proposed Methodology}
\label{sec:methodology}

\subsection{System Model and Architecture Overview}
We consider the case of a smart city environmental monitoring system, which consists of a large number of geographically dispersed sensors in the city. This system is designed to operate under severe energy constraints assuming that continuous communication and sensing are impractical for long term deployments.

As illustrated in Fig.~\ref{fig:system_diagram}, the proposed framework is based on a three-layer horizontal architecture. From left to right, we can see a Sensor Layer, an Edge Intelligence Layer, and a Cloud Layer. The key design is the hierarchical allocation of responsibilities. The Sensor Layer are used for local inference and summarization of the environmental dynamics; the Edge Intelligence Layer will make decisions based on contextual information to facilitate sensor activation; the Cloud Layer is used to provide passive data aggregation, long-term storage, and visualization.

\subsection{Sensor Layer (TinyML-Enabled Edge Devices)}
We let $S = \{s_1, s_2, \dots, s_N\}$ denote the set of environmental sensor nodes deployed across the city. Each sensor node is equipped with:
\begin{itemize}
    \item One or more environmental sensors (e.g., air quality, noise, temperature),
    \item A low-power microcontroller capable of running TinyML inference,
    \item A limited energy budget determined by battery capacity or energy harvesting.
\end{itemize}
Rather than continuously transmitting raw sensor readings, each node uses a TinyML model to generate a brief feedback signal summarizing the local environmental dynamics. This feedback can for example comprise of:
\begin{itemize}
    \item A change or anomaly indicator,
    \item Signal variability or confidence score,
    \item A coarse event likelihood estimate.
\end{itemize}
We denote the summarized feedback of sensor $s$ at time $t$ as $\hat{I}_s(t)$. This design will minimize the communication overhead and allows sensors to remain in a dormant state unless being explicitly activated by the Edge Intelligence Layer.

\subsection{Edge Intelligence Layer (System Decision Core)}
The Edge Intelligence Layer acts as the core of the decision process which manages a group of nearby sensors (e.g., within a geographic cluster). It operates in discrete monitoring intervals $t = 1, 2, \dots$, and follows an observe--decide--update cycle. At each monitoring interval $t$, the Edge Intelligence Layer observes a context vector $C_t$, which may include:
\begin{itemize}
    \item Time of day,
    \item Geographic zone or cluster identifier,
    \item Recent aggregated environmental trends,
    \item Remaining energy levels of sensor nodes.
\end{itemize}

The Edge Intelligence Layer will take this context vector as input to determine which subset of sensors should be activated for full sensing and transmission in the current interval. For high scalability, the sensors are logically organized into clusters which ensures the activation decision can be made hierarchically. The high-level decisions will identify regions of interest, and the local decisions will select individual sensors in a cluster.

\subsection{Cloud Layer}
The Cloud Layer receives compressed summary information and aggregated metrics in a passive manner from the Edge Intelligence Layer. It can support long-term storage, visualization dashboards, and reporting for city operators. Note that in our framework the cloud does not take part in the sensing decisions, which will preserve the system's low latency and energy efficiency.

\subsection{Energy-Aware Utility Model}
Every sensor $s$ has an associated estimated energy cost $E_s$ which represents the overall cost of sensing and data transmission. The framework also maintains a utility estimate $U_s$ for every sensor to balance the tradeoff between monitoring effectiveness and energy efficiency.

The sensing reward is defined as follows:
\begin{equation}
R_s = \alpha \cdot I_s - \beta \cdot E_s,
\label{eq:reward}
\end{equation}
where:
\begin{itemize}
    \item $I_s$ denotes the information gain associated with activating sensor $s$,
    \item $E_s$ is the corresponding energy cost,
    \item $\alpha$ and $\beta$ are between factors to strike a balance between the quality of monitoring and energy comsumption.
\end{itemize}
Rather than estimating $I_s$ directly, the framework uses the feedback provided by TinyML, denoted as $\hat{I}_s$ as a lightweight proxy, which makes it more practical for edge deployment.

\begin{algorithm}[t!]
\caption{Adaptive Edge-Based Sensor Activation}
\label{alg:adaptive_activation}
\begin{algorithmic}[1]
\REQUIRE Sensor set $S$, energy budget $B$, learning rate $\eta$, energy costs $\{E_s\}_{s\in S}$
\STATE Initialize utility estimates $U_s$ for all $s \in S$
\FOR{each monitoring interval $t$}
\STATE Observe current context $C_t$
\FOR{each sensor $s \in S$}
\STATE Compute $\text{Score}_s \leftarrow U_s / E_s$
\ENDFOR
\STATE Select a subset $S_t \subseteq S$ of top-$K$ sensors maximizing $\text{Score}_s$ subject to budget $B$
\STATE Activate sensors in $S_t$ and collect feedback $\hat{I}_s$ for all $s \in S_t$
\FOR{each sensor $s \in S_t$}
\STATE Update $U_s \leftarrow (1-\eta)\cdot U_s + \eta \cdot \hat{I}_s$
\ENDFOR
\ENDFOR
\end{algorithmic}
\end{algorithm}

\subsection{Adaptive Sensor Activation Algorithm}
The Adaptive Sensor Activation Algorithm is only executed in the Edge Intelligence Layer and follows a discrete-time online decision process. It is specifically designed to be lightweight, transparent, and computationally inexpensive, which ensures the algorithm’s appropriateness for real-time application in devices that manage extensive sets of resource-constrained sensors.

In each monitoring interval $t = 1, 2, \dots$, the algorithm consists of three stages: scoring, constrained selection, and online update.

\subsubsection{Sensor Scoring and Ranking}
At the beginning of each monitoring period $t$, the edge coordinator has a utility estimate $U_s(t)$ for each sensor $s \in S$, reflecting the historical usefulness of the sensor in detecting useful environmental changes.

To incorporate energy awareness, each sensor $s$ will be associated with an activation cost estimate $E_s$ which summarizes the related energy costs. The edge coordinator will calculate a normalized selection score for every sensor:

\begin{equation}
\text{Score}_s(t) = \frac{U_s(t)}{E_s}.
\label{eq:score}
\end{equation}
This score measures the expected information contribution per unit energy, which in turn enables fair comparison of different sensors that have different energy profiles. Sensors that have higher scores are considered to be more cost-effective to be activated.

\subsubsection{Top-K Selection Under Energy Budget}
For a given energy budget $B$ in the current interval, the edge coordinator identifies which sensors are to be activated. More specifically, the sensors are ranked in descending order of $\text{Score}_s(t)$, and the top-$K$ sensors are selected such that:
\begin{equation}
\sum_{s \in S_t} E_s \le B,
\end{equation}
where $S_t \subseteq S$ denotes the selected sensor subset at time $t$.

The value of $K$ is not fixed in advance but implicitly decided by the available energy budget $B$ and the energy cost of each sensor node $E_s$. Such mechanism can adaptively control the number of active sensors according to the constraints.

Only sensors in the set $S_t$ are eligible to receive activation commands and are active in terms of sensing and transmission during the interval. Other sensors are in a low power or sleep state, which saves power.

\subsubsection{Feedback Collection and Online Utility Update}
After being activated, each sensor $s \in S_t$ sends a concise feedback signal $\hat{I}_s(t)$, which is a result of the sensor’s internal TinyML inference. This feedback signal can be regarded as a proxy of the informational value of the sensor readings.

Subsequently, the edge coordinator updates the estimate of the utility of each activated sensor by using an exponential moving average update rule:
\begin{equation}
U_s(t+1) \leftarrow (1-\eta)\cdot U_s(t) + \eta\cdot \hat{I}_s(t),
\label{eq:update}
\end{equation}
where $\eta \in (0,1)$ is a learning rate that determines the rate of adaptation, and sensors that are not activated in interval $t$ maintain their previously computed utilities.

This update rule allows the framework to dynamically adjust to new changes in the urban environment, eliminating the need to retrain the model or store large amounts of data.

\subsubsection{Algorithm Summary and Complexity}
Algorithm~\ref{alg:adaptive_activation} describes the overall adaptive sensor activation process, including initialization, interval-wise scoring, top-K selection subject to energy constraints, activation, and finally, update to the utility.

We can observe that the time complexity of the algorithm is linearly dependent on the number of sensors in each edge cluster, denoted by ($O(N)$), and dominated by score computation and ranking. Hence, the algorithm is highly efficient in terms of real-time execution at the edge which facilitates large-scale sensor networks.

\section{Results and Discussion}
\label{sec:results}

\begin{figure*}[t!]
\centering
\begin{subfigure}[b]{0.3\textwidth}
    \centering
    \resizebox{\linewidth}{!}{%
    \begin{tikzpicture}
      \draw[thick, <->] (0,3.5) node[above, font=\small] {\textbf{Energy (mAh)}} -- (0,0) -- (6.4,0);
      \foreach \y/\label in {0.75/30, 1.5/60, 2.25/90, 3.0/120} {
         \draw[gray!30, dashed] (0,\y) -- (6.4,\y);
         \node[left, font=\footnotesize] at (0,\y) {\label};
      }
      \fill[customGray] (0.4,0) rectangle (1.2, 3.0);
      \node[above, font=\footnotesize] at (0.8,3.0) {\textbf{120}};
      \node[below, align=center, font=\footnotesize] at (0.8,0) {Static};

      \fill[customBlue] (1.8,0) rectangle (2.6, 2.375);
      \node[above, font=\footnotesize] at (2.2,2.375) {\textbf{95}};
      \node[below, align=center, font=\footnotesize] at (2.2,0) {Periodic};

      \fill[customOrange] (3.2,0) rectangle (4.0, 1.9);
      \node[above, font=\footnotesize] at (3.6,1.9) {\textbf{76}};
      \node[below, align=center, font=\footnotesize] at (3.6,0) {\rev{UCB}};

      \fill[customGreen] (4.6,0) rectangle (5.4, 1.75);
      \node[above, font=\footnotesize] at (5.0,1.75) {\textbf{70}};
      \node[below, align=center, font=\footnotesize] at (5.0,0) {\textbf{Proposed}};
    \end{tikzpicture}
    }
    \caption{Daily Energy Consumption}
    \label{fig:energy_sub}
\end{subfigure}
\hfill
\begin{subfigure}[b]{0.3\textwidth}
    \centering
    \resizebox{\linewidth}{!}{%
    \begin{tikzpicture}
      \draw[thick, <->] (0,3.5) node[above, font=\small] {\textbf{Detection Rate (\%)}} -- (0,0) -- (6.4,0);
      \foreach \y/\label in {0.75/25, 1.5/50, 2.25/75, 3.0/100} {
         \draw[gray!30, dashed] (0,\y) -- (6.4,\y);
         \node[left, font=\footnotesize] at (0,\y) {\label};
      }
      \fill[customGray] (0.4,0) rectangle (1.2, 2.34);
      \node[above, font=\footnotesize] at (0.8,2.34) {78\%};
      \node[below, align=center, font=\footnotesize] at (0.8,0) {Static};

      \fill[customBlue] (1.8,0) rectangle (2.6, 2.46);
      \node[above, font=\footnotesize] at (2.2,2.46) {82\%};
      \node[below, align=center, font=\footnotesize] at (2.2,0) {Periodic};

      \fill[customOrange] (3.2,0) rectangle (4.0, 2.67);
      \node[above, font=\footnotesize] at (3.6,2.67) {\textbf{89\%}};
      \node[below, align=center, font=\footnotesize] at (3.6,0) {\rev{UCB}};

      \fill[customGreen] (4.6,0) rectangle (5.4, 2.73);
      \node[above, font=\footnotesize] at (5.0,2.73) {\textbf{91\%}};
      \node[below, align=center, font=\footnotesize] at (5.0,0) {\textbf{Proposed}};
    \end{tikzpicture}
    }
    \caption{Event Detection Rate}
    \label{fig:coverage_sub}
\end{subfigure}
\hfill
\begin{subfigure}[b]{0.3\textwidth}
    \centering
    \resizebox{\linewidth}{!}{%
    \begin{tikzpicture}
      \draw[thick, <->] (0,3.5) node[above, font=\small] {\textbf{Lifetime (Days)}} -- (0,0) -- (6.4,0);
      \foreach \y/\label in {1.0/100, 2.0/200, 3.0/300} {
         \draw[gray!30, dashed] (0,\y) -- (6.4,\y);
         \node[left, font=\footnotesize] at (0,\y) {\label};
      }
      \fill[customGray] (0.4,0) rectangle (1.2, 1.8);
      \node[above, font=\footnotesize] at (0.8,1.8) {180};
      \node[below, align=center, font=\footnotesize] at (0.8,0) {Static};

      \fill[customBlue] (1.8,0) rectangle (2.6, 2.27);
      \node[above, font=\footnotesize] at (2.2,2.27) {227};
      \node[below, align=center, font=\footnotesize] at (2.2,0) {Periodic};

      \fill[customOrange] (3.2,0) rectangle (4.0, 2.84);
      \node[above, font=\footnotesize] at (3.6,2.84) {\textbf{284}};
      \node[below, align=center, font=\footnotesize] at (3.6,0) {\rev{UCB}};

      \fill[customGreen] (4.6,0) rectangle (5.4, 3.09);
      \node[above, font=\footnotesize] at (5.0,3.09) {\textbf{309}};
      \node[below, align=center, font=\footnotesize] at (5.0,0) {\textbf{Proposed}};
    \end{tikzpicture}
    }
    \caption{Estimated Network Lifetime}
    \label{fig:lifetime_sub}
\end{subfigure}
\caption{\rev{Performance evaluation results for four sensing strategies: static sensing, periodic sensing, UCB-based sensing, and the proposed framework.} (a) Average daily energy consumption, (b) High-pollution event detection rate, and (c) estimated sensor network lifetime. The proposed framework outperforms baseline strategies across all metrics.}
\label{fig:results_all}
\end{figure*}
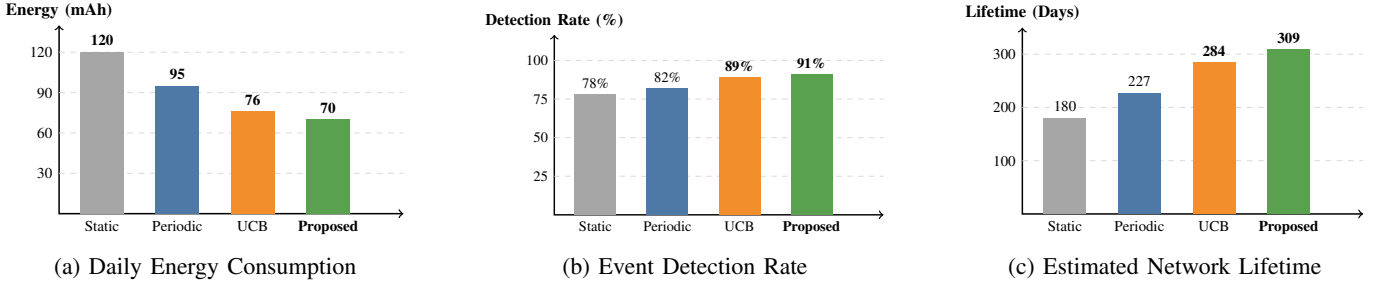

\subsection{Simulation Setup}
The proposed framework was tested within a trace-driven city-scale simulation based on the UK Air Quality Reanalysis (AQREAN) dataset \cite{b22}. AQREAN provides hourly gridded air-quality fields on a 0.1$^\circ$ latitude-longitude grid. In our simulation, 20 grid cells were mapped to 20 geographical zones, and the traces of PM$_{2.5}$, PM$_{10}$, CO, NO$_2$, O$_3$, and SO$_2$ were used to drive zone-level dynamics. The hourly data were linearly interpolated to 15-minute intervals, which yields 2880 decision rounds over 30 days. Each zone contains 50 heterogeneous virtual sensor nodes, for 1000 nodes in total. Node readings are generated from the corresponding AQREAN trace with sensor noise, heterogeneous energy costs, and battery levels, and the episodic high-pollution events are injected to test detection performance. The context includes zone, time-of-day, recent pollutant levels, short-term trends, and sensor energy state. We consider three comparison strategies for evaluation:
\begin{itemize}
    \item \textbf{Static Sensing:} All sensors remain active continuously.
    \item \textbf{Periodic Sensing:} Sensors are activated at fixed time intervals.
    \item \rev{\textbf{UCB-Based Adaptive Sensing:} Sensors are selected based on an upper-confidence-bound policy according to the historical sensing reward and energy cost, without any context-aware mechanism.}
\end{itemize}
In the following analysis, we focus on sensor-node energy consumption. The computation cost of the Edge Intelligence Layer is assumed negligible compared with sensing and communication cost.

\subsection{Energy Consumption Analysis}
\begin{table}[h]
\caption{Energy Consumption Comparison}
\label{tab:energy}
\centering
\begin{tabular}{lcc}
\toprule
\textbf{Method} & \textbf{Avg Energy (mAh/day)} & \textbf{Reduction} \\
\midrule
Static Sensing & 120 & -- \\
Periodic Sensing & 95 & 21\% \\
\rev{UCB-Based Sensing} & \rev{76} & \rev{36.7\%} \\
\textbf{Proposed Framework} & \textbf{70} & \textbf{42\%} \\
\bottomrule
\end{tabular}
\end{table}

Figure \ref{fig:results_all}(a) and Table \ref{tab:energy} compare the average daily energy consumption of the \rev{four} sensing strategies. Static sensing consumes 120 mAh/day, while periodic sensing reduces this to 95 mAh/day. \rev{The UCB-based sensing strategy further reduces daily energy consumption to 76 mAh/day by adaptively prioritizing sensors with higher confidence-adjusted utility.}

The proposed framework shows the lowest amount of daily energy consumption at 70 mAh/day, which reflects a 42\% reduction when compared to the case of static sensing. \rev{It also outperforms the UCB baseline because the proposed adaptive sensor activation algorithm selects only a limited number of sensors at each interval since it uses historical utility, spatiotemporal context and TinyML-derived local feedback.}

\subsection{Environmental Monitoring Coverage}
\begin{table}[h]
\caption{Detection Coverage}
\label{tab:coverage}
\centering
\begin{tabular}{lcc}
\toprule
\textbf{Method} & \textbf{Coverage Rate (\%)} & \textbf{Improvement} \\
\midrule
Static Sensing & 78 & -- \\
Periodic Sensing & 82 & +5.1\% \\
\rev{UCB-Based Sensing} & \rev{89} & \rev{+14.1\%} \\
\textbf{Proposed Framework} & \textbf{91} & \textbf{+16.7\%} \\
\bottomrule
\end{tabular}
\end{table}

Fig.~\ref{fig:results_all}(b) and Table~\ref{tab:coverage} show the detection rate of high-pollution events for the \rev{four} strategies. Although our proposed framework uses fewer sensors, it achieves the highest detection rate, 91\%. This is higher than the static, periodic, and \rev{UCB-based} strategies, which achieve 78\%, 82\%, and \rev{89\%}, respectively. \rev{The UCB baseline already improves coverage over periodic sensing, confirming that online learning is a stronger benchmark than fixed scheduling alone.} The improvement can be attributed to the fact that the algorithm adapts to variable regions in the environment rather than using uniform sampling.

\rev{The comparison with UCB further shows that context-aware sensing can improve monitoring effectiveness beyond generic confidence-based exploration while still reducing the overall sensing activity.}

\subsection{Sensor Network Lifetime}
\begin{table}[h]
\caption{Network Lifetime}
\label{tab:lifetime}
\centering
\begin{tabular}{lcc}
\toprule
\textbf{Method} & \textbf{Sensor Lifetime (Days)} & \textbf{Extension} \\
\midrule
Static Sensing & 180 & -- \\
Periodic Sensing & 227 & +26.1\% \\
\rev{UCB-Based Sensing} & \rev{284} & \rev{+57.8\%} \\
\textbf{Proposed Framework} & \textbf{309} & \textbf{+71.7\%} \\
\bottomrule
\end{tabular}
\end{table}

The sensor network lifetime is directly determined from the daily energy consumption under the assumption of identical nominal battery capacity. Based on the static sensing scenario, the battery capacity is estimated as $C_{\text{bat}} = 120 \times 180 = 21600$ mAh. \rev{Using this capacity, the lifespan of the periodic sensing strategy, UCB-based sensing strategy, and the proposed strategy are estimated as $21600/95 \approx 227$ days, $21600/76 \approx 284$ days, and $21600/70 \approx 309$ days, respectively.}

As shown in Fig.~\ref{fig:results_all}(c) and Table~\ref{tab:lifetime}, the proposed framework significantly improves sensor network lifetime by avoiding unnecessary activation and allowing non-selected nodes to remain in low-power state for longer. The UCB baseline also extends lifetime substantially compared with static and periodic sensing, but the proposed method remains superior because it better aligns activation decisions with context and current environmental informativeness.

\subsection{Discussion}
From the obtained results we can see that, when combined with adaptive AI methodologies, edge intelligence can balance energy efficiency, monitoring accuracy, and network lifetime. Different from static or periodic sensing paradigms, our proposed framework ensures more effective alignment of sensing decisions with environmental relevance and resource constraints. The additional UCB comparison also strengthens the novelty justification of this work: while UCB-based adaptive sensing is already more competitive than fixed scheduling, the proposed framework consistently performs better because it integrates TinyML-enabled local feedback, explicit spatiotemporal context-awareness, and hierarchical edge coordination.

\subsection{Limitations}
Although the results are promising, several practical issues remain. First, the simulations assume predominantly stable wireless links, whereas city-canyon environments may introduce dropouts that delay activation commands. Second, the analysis assumes a uniform degradation rate across sensor types, although low-cost devices may fail unpredictably for reasons unrelated to battery life. Finally, manual battery replacement cost is not included in the current model.

\section{Conclusion}
\label{sec:conclusion}
Environmental monitoring plays a key role in enabling sustainable and resilient smart cities. However, extensive sensor networks often face challenges related to limited power supplies and inefficient static sensing strategies. This paper proposes an AI-based framework for energy-efficient environmental monitoring in smart cities using edge computing to address these challenges.

By employing TinyML-based edge devices and an energy-aware decision process, the framework selects the most informative sensors in real time based on environmental conditions and available resources. To support scalability, a hierarchical Edge Intelligence architecture is introduced. Simulations show that energy consumption is reduced considerably and sensor lifetime is increased while maintaining robust environmental monitoring capability compared with static sensing, periodic sensing, and a UCB-based adaptive baseline.

The results indicate that edge intelligence and adaptive AI provide a practical approach to sustainable sensing infrastructure in smart cities. Future work will focus on practical pilots, heterogeneous sensors, and advanced learning mechanisms to improve robustness and adaptability in complex smart-city environments.


\end{document}